\documentclass[aps,showpacs,floatfix,twocolumn,superscriptaddress]{revtex4}

\usepackage{graphicx} 
\usepackage{dcolumn} 
\usepackage{amsmath} 

\newcommand{\sfrac}[2]{\mbox{\footnotesize $\displaystyle \frac{#1}{#2}$}} 
\newcommand{\ssize}[1]{\mbox{\footnotesize $\displaystyle {#1}$}}
\newcommand{\ihsp}{\hspace*{\fill} }

\newcommand{\be}{\begin{equation}}
\newcommand{\ee}{\end{equation}}
\newcommand{\bea}{\begin{eqnarray}}
\newcommand{\eea}{\end{eqnarray}}

\newcommand{\WT}{Ward-Takahashi }

\newcommand{\Eq}[1]{Eq.~(\ref{#1})}
\newcommand{\Eqs}[1]{Eqs.~(\ref{#1})}
\newcommand{\Fig}[1]{Fig.~{\ref{#1}}}

\newcommand{\Ref}[1]{Ref.~\cite{#1}}
\newcommand{\Refs}[1]{Refs.~\cite{#1}}

\def\slr#1{\setbox0=\hbox{$#1$}           
   \dimen0=\wd0                                 
   \setbox1=\hbox{/} \dimen1=\wd1               
   \ifdim\dimen0>\dimen1                        
      \rlap{\hbox to \dimen0{\hfil/\hfil}}      
      #1                                        
   \else                                        
      \rlap{\hbox to \dimen1{\hfil$#1$\hfil}}   
      /                                         
   \fi}

\begin{document}
\title{Quark-gluon vertex model and lattice-QCD data}
\author{M.~S.~Bhagwat}
\affiliation{Center for Nuclear Research, Department of Physics,  Kent State 
University, Kent OH 44242, USA}
\author{P.~C.~Tandy}
\affiliation{Center for Nuclear Research, Department of Physics,  Kent State 
University, Kent OH 44242, USA}
\date{\today}

\begin{abstract}
A model for the dressed quark-gluon vertex, at zero gluon momentum, is 
formed from a nonperturbative extension of the two Feynman diagrams that 
contribute at 1-loop in perturbation theory.  The required
input is an existing ladder-rainbow model Bethe-Salpeter kernel from an approach 
based on the Dyson-Schwinger equations; no new
parameters are introduced.   The model includes an Ansatz for the 
triple-gluon vertex.   Two of the three vertex amplitudes from the 
model provide a point-wise description of the recent quenched lattice-QCD
data.  An estimate of the effects of quenching is made.
\end{abstract}


%


\preprint{Kent State U preprint no. KSUCNR-204-03}
\maketitle
%
\section{\label{sec:intro} Introduction}

A great deal of progress in the QCD modeling of hadron physics has been 
achieved through the use of the ladder-rainbow truncation of the Dyson-Schwinger
equations (DSEs).   For two recent reviews, see \Refs{Maris:2003vk} and 
\cite{Alkofer:2000wg}.  Apart from 1-loop renormalization group improvement,
this truncation is built upon a bare quark-gluon vertex. 
Recent investigations with simple dressed vertex models have indicated 
that material contributions to a number of observables are possible with
a better understanding of the infrared structure of the vertex.  These
diverse model indications include an enhancement in the quark 
condensate~\cite{Fischer:2003rp,Bhagwat:2004hn}, an increase of about 300~MeV
in the $b_1/h_1$ axial vector meson mass~\cite{Watson:2004kd},
and about 200~MeV of attraction in the $\rho/\omega$ vector meson mass.

In the absence of well-constrained 
nonperturbative models for the vertex, it has often been assumed that
a reasonable beginning is the (Abelian) Ball-Chiu Ansatz~\cite{Ball:1980ay} 
times the appropriate color matrix.   An example is provided by 
the recent results from a truncation of the gluon-ghost-quark DSEs where
this vertex dressing contributes materially to a reasonable quark condensate 
value~\cite{Fischer:2003rp}.   However, there is no known way to develop 
a Bethe-Salpeter (BSE) kernel that is 
dynamically matched to a quark self-energy defined in terms of such
a phenomenological dressed vertex in the sense
that chiral symmetry is preserved through the axial-vector \WT identity.
The latter implementation of chiral symmetry guarantees the Goldstone boson 
nature of the flavor non-singlet 
pseudoscalars independently of model details~\cite{Maris:1998hd}.
There is a known  constructive scheme~\cite{Bender:1996bb} that defines 
a diagrammatic expansion of 
the BSE kernel corresponding to any diagrammatic expansion of the quark 
self-energy such that the axial-vector \WT  identity is preserved.   
For this reason, recent nonperturbative vertex models have employed simple 
diagrammatic 
representations~\cite{Bhagwat:2004hn,Watson:2004kd,FischerAdel04}.

It is only recently that lattice-QCD has begun to provide information
on the infrared structure of the dressed quark-gluon 
vertex~\cite{Skullerud:2003qu}.   In this work we generate a model dressed
vertex, for zero gluon momentum, based on an Ansatz for non-perturbative 
extensions of the only two diagrams that contribute at 1-loop order in 
perturbation theory.   An existing ladder-rainbow model kernel is the
only required input.  We compare to the recent lattice-QCD data without 
parameter adjustment.
 
In Section~\ref{sec:pert} we recall the vertex to 1-loop 
in perturbation theory and point out the structure and properties that
are used to suggest the Ansatz for non-perturbative extension.   The 
non-perturbative extension is described in Section~\ref{sec:nonpert} and
the results are presented and discussed in Section~\ref{sec:results}.


\section{\label{sec:pert} One-loop perturbative vertex}

\begin{figure}[h]
\vspace*{-8mm}
\centering{\
\includegraphics[width=40mm]{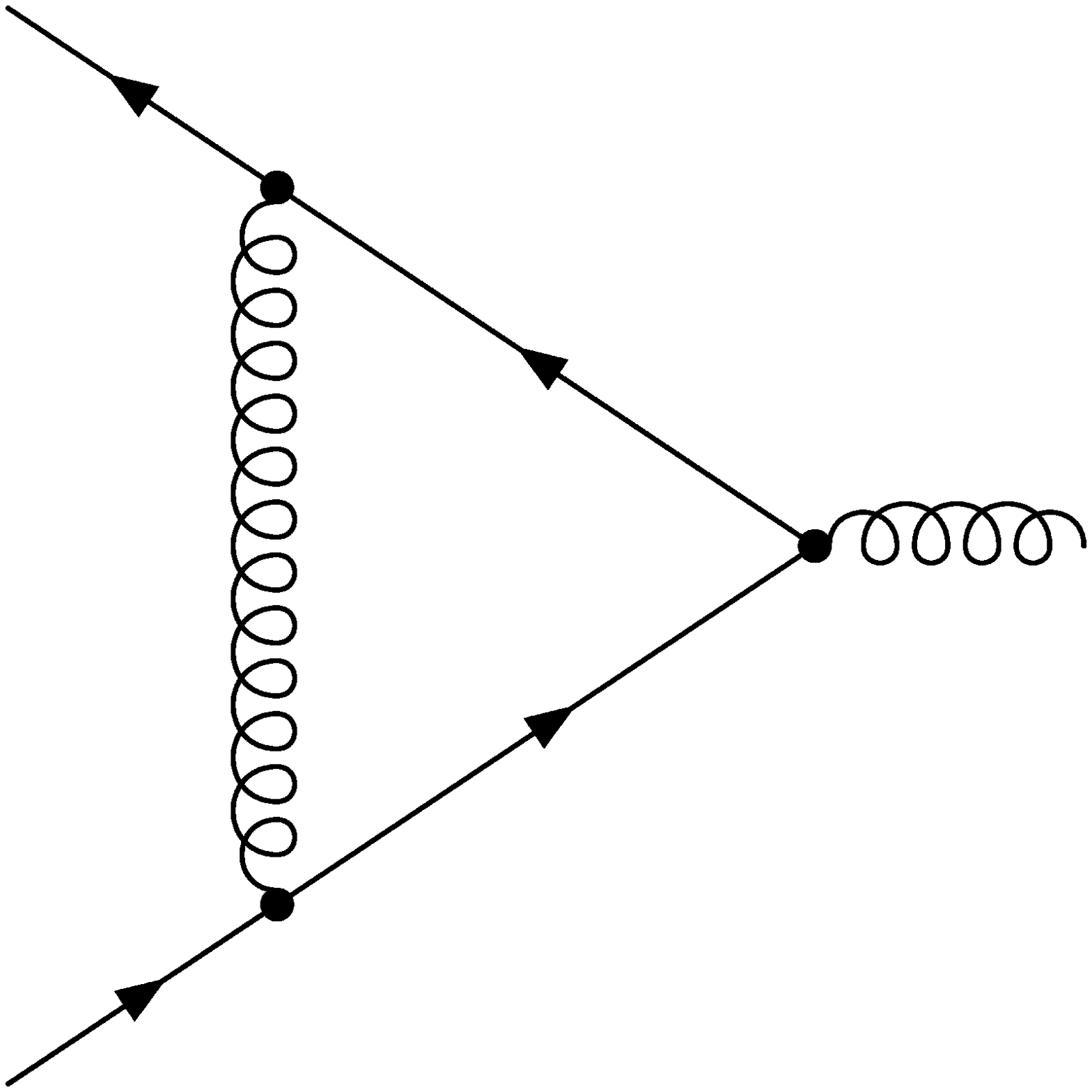}\ihsp
\includegraphics[width=40mm]{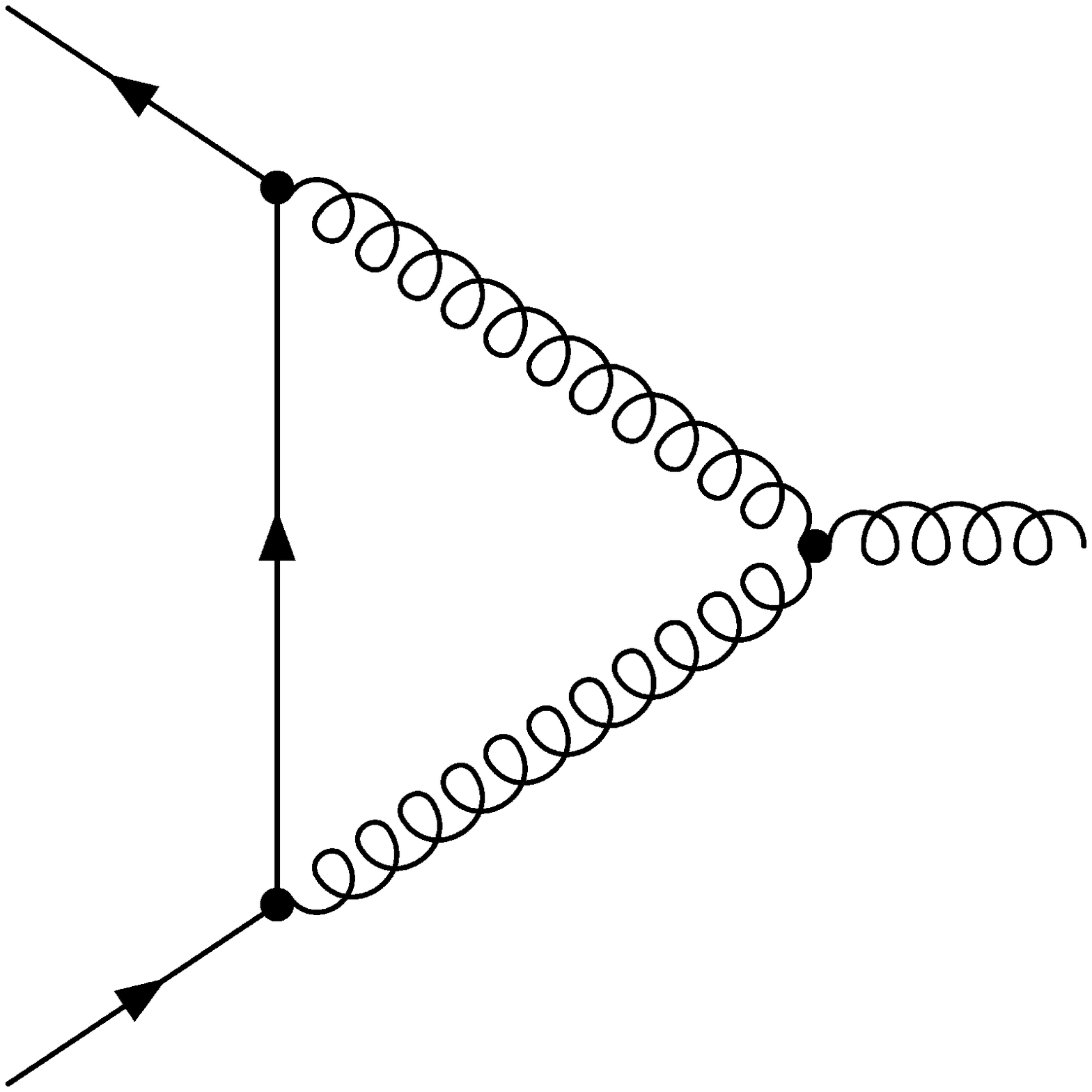} }
\vspace*{-10mm}
\caption{\label{fig:2vertdiags} The quark-gluon vertex at one loop. 
The left diagram labelled A is the Abelian-like term 
$\Gamma_\sigma^{\rm A}$, and the right diagram labelled NA is
the non-Abelian term $\Gamma_\sigma^{\rm NA}$. }
\vspace*{-20mm}
\centerline{\ihsp A \ihsp \hspace*{20mm} NA \ihsp}
\vspace*{20mm}
\end{figure}

We denote the  dressed-quark-gluon vertex for gluon momentum $k$ and quark 
momentum $p$ by \mbox{$ig\, t^c\,\Gamma_\sigma(p+k,p)$}, where 
\mbox{$t^c = \lambda^c/2$} and $\lambda^c$ is an SU(3) color matrix.  
Through ${\cal O}(g^2)$, i.e., to 1-loop, the amplitude $\Gamma_\sigma$ is given,
in terms of \Fig{fig:2vertdiags}, by$^{\rm\footnotemark[1]}$
\footnotetext[1]{We employ Landau gauge and a Euclidean metric, with: 
$\{\gamma_\mu,\gamma_\nu\} = 2\delta_{\mu\nu}$; $\gamma_\mu^\dagger = 
\gamma_\mu$; and $a \cdot b = \sum_{i=1}^4 a_i b_i$.}   
\begin{equation}
\label{vert1loop}
\Gamma_\sigma(p+k,p)  =  Z_{\rm 1F}\,\gamma_\sigma + 
\Gamma_\sigma^{\rm A}(p+k,p) +  \Gamma_\sigma^{\rm NA}(p+k,p)+ \ldots~~,
\end{equation}
with
\begin{eqnarray} 
\Gamma_\sigma^{\rm A}(p+k,p) &=& -(\ssize{C_{\rm F}}-\sfrac{C_{\rm A}}{2})\,
\int_q^\Lambda\! g^2 D_{\mu\nu}(p-q)\gamma_\mu \nonumber\\
&& \times S_0(q+k)\, \gamma_\sigma\, S_0(q) \gamma_\nu~~~, 
\label{VertA} 
\end{eqnarray}
and
\begin{eqnarray}
\Gamma_\sigma^{\rm NA}(p+k,p) &=& - \sfrac{C_{\rm A}}{2} \int_q^\Lambda\! 
g^2\,\gamma_\mu S_0(p-q) \gamma_\nu\, D_{\mu \mu^\prime}(q+k)\,
\nonumber\\
& & \times  
i\Gamma^{3g}_{\mu^\prime \nu^\prime \sigma}(q+k,q)\,
                                        D_{\nu^\prime \nu}(q)~~~,
\label{VertNA}
\end{eqnarray} 
where \mbox{$\int^\Lambda_q =$} \mbox{$\int^\Lambda d^4 q/(2\pi)^4$} 
denotes a loop integral  regularized in a translationally-invariant manner
at mass-scale $\Lambda$.   Here $Z_{\rm 1F}(\mu^2,\Lambda^2)$ is the 
vertex renormalization constant to ensure \mbox{$\Gamma_\sigma = \gamma_\sigma$}
at renormalization scale $\mu$.   The following quantities are bare:
the three-gluon vertex
$ ig\,f^{\rm abc}\,\Gamma^{3g}_{\mu \nu \sigma}(q+k,q)$, the quark propagator 
$S_0(p)$, and the gluon propagator 
\mbox{$D_{\mu \nu}(q)= T_{\mu \nu}(q) D_0(q^2)$}, where  $T_{\mu \nu}(q)$ 
is the transverse projector.  
The next order terms in \Eq{vert1loop} are ${\cal O}(g^3)$: the contribution 
involving the four-gluon vertex, and  ${\cal O}(g^4)$: contributions 
from crossed-box and two-rung gluon ladder diagrams, and 1-loop dressing
of the triple-gluon vertex, etc.  

The color factors in \Eqs{VertA} and (\ref{VertNA}), given by
\begin{eqnarray}
t^a\, t^b\, t^a &=& (\ssize{C_{\rm F}}-\sfrac{C_{\rm A}}{2})\, t^b
             = - \sfrac{1}{2 N_c}\, t^b \nonumber\\
t^a\, f^{abc}\, t^b &=& \sfrac{C_{\rm A}}{2} i\, t^c = 
                                     \sfrac{N_c}{2} i\, t^c~~~, 
\label{colorfacts}
\end{eqnarray}
reveal two important considerations.  The color factor of the
(Abelian-like) term $\Gamma_\sigma^{\rm A}$ would be given by
\mbox{$t^a\,t^a = \,\ssize{C_{\rm F}} = \ssize{(N_c^2-1)}/ \ssize{2 N_c}$}
for the strong dressing of the photon-quark vertex, i.e., in the color
singlet channel.  The octet $\Gamma_\sigma^{\rm A}$ is of opposite sign and is 
suppressed 
by a factor $1/(N_c^2-1)$: single gluon exchange between a quark and antiquark 
has relatively weak repulsion in the color-octet channel, compared to strong
attraction in the color-singlet channel.  Net attraction for the gluon vertex
(at least to this order)  is provided by the non-Abelian 
$\Gamma_\sigma^{\rm NA}$ term, which involves the three-gluon vertex:
the color factor is amplified by $-N_c^2$ over the $\Gamma_\sigma^{\rm A}$
term.

The specific form of the bare triple-gluon vertex is conveniently expressed  
in terms of three momenta  $p_1=q+k$, $p_2=-q$ and $p_3=-k$, that are outgoing.
Thus with \mbox{$\Gamma^{3g}_{\mu \nu \sigma}(q+k,q) \equiv 
\tilde{\Gamma}^{3g}_{\mu \nu \sigma}(p_1,p_2,p_3)$}, we have
\bea
\tilde{\Gamma}^{3g}_{\mu \nu \sigma}(p_1,p_2,p_3) &=& 
- \big\{ (p_1-p_2)_\sigma\, \delta_{\mu\nu}
+ (p_2-p_3)_\mu\, \delta_{\nu\sigma} \nonumber\\
&& {} + (p_3-p_1)_\nu\, \delta_{\sigma\mu}\big\}~~~,
\label{bare3Gvert}
\eea
and the complete vertex is symmetric under permutations of all gluon 
coordinates.  In Landau gauge  $\Gamma^{3g}_{\mu \nu \sigma}$ obeys the 
Slavnov-Taylor identity 
\be
k_\sigma\,\Gamma^{3g}_{\mu \nu \sigma}(q+k,q) = 
  D_0^{-1}(q)\, T_{\mu \nu}(q) -  D_0^{-1}(q+k)\, 
T_{\mu \nu}(q+k)~.\\
\label{3GSTI}
\ee
The nonperturbative model of Section \ref{sec:nonpert} addresses the $k=0$
case and makes an extension of the bare result 
\be
\Gamma^{3g}_{\mu \nu \sigma}(q,q) = 
-\,\sfrac{\partial}{\partial q_\sigma}
                      D_0^{-1}(q)\, T_{\mu \nu}(q)~~~,
\label{3GSTI0}
\ee
which allows the amplitude for the non-Abelian diagram at $k=0$ to 
take the form 
\bea
\Gamma_\sigma^{\rm NA}(p,p) &=& - i\,\sfrac{C_{\rm A}}{2} \int_q^\Lambda\! 
\gamma_\mu S_0(p-q) \gamma_\nu\, \nonumber\\
&& \times \left\{\sfrac{\partial}{\partial q_\sigma} g^2\,D_0(q^2)
\right\}  T_{\mu \nu}(q)~~~.
\label{VertNA0}
\eea
It is easy to verify that the Abelian diagram gives 
\be
\Gamma_\sigma^{\rm A}(p,p) = -i [1-\sfrac{C_{\rm A}}{2}\, 
                                   \ssize{C_{\rm F}^{-1}}]\,
                 \sfrac{\partial}{\partial p_\sigma} \Sigma^{(1)}(p)~~~,  
\label{VertAderivfm}     
\ee
in terms of the 1-loop self-energy.

The dressing provided by the combination 
\mbox{$\Gamma_\sigma^{\rm A} + \Gamma_\sigma^{\rm NA}$}  
yields a vertex that satisfies the Slavnov-Taylor identity (STI) through 
${\cal O}(g^2)$~\cite{Davydychev:2000rt}.  This identity expresses the divergence 
of the vertex in terms of the bare and 1-loop contributions to three objects:
$S(p)^{-1}$, the ghost propagator dressing function, and the ghost-quark 
scattering amplitude.  The 1-loop $S(p)^{-1}$ part of this relation is 
generated partly from $\Gamma_\sigma^{\rm A}$ (with a weak repulsive color
strength) and partly from $\Gamma_\sigma^{\rm NA}$ (with the complementary
strongly attractive color strength).  The $\Gamma_\sigma^{\rm NA}$
term also provides the explicitly non-Abelian terms of the ${\cal O}(g^2)$ STI.

\section{\label{sec:nonpert} Nonperturbative vertex model}

The general nonperturbative vertex at $k=0$ has a representation in terms of
three invariant amplitudes; here we choose 
\be
\Gamma_\sigma(p,p) = \gamma_\mu \lambda_1(p^2)
                     - 4 p_\mu\, \gamma \cdot p\, \lambda_2(p^2)
                     - i 2 p_\mu\, \lambda_3(p^2)~~~.\\
\label{lambda_def}
\ee
since the lattice-QCD data~\cite{Skullerud:2003qu} is provided in terms of 
these $\lambda_i(p^2)$ amplitudes.
A useful comparison is the corresponding vertex in an Abelian theory 
like QED; it is given by the Ward identity \mbox{$\Gamma_\sigma^{WI}(p,p) =$} 
\mbox{$ -i \partial S^{-1}(p)/\partial p_\sigma$} in terms of the exact 
propagator $S^{-1}(p)$.  With \mbox{$S^{-1}(p) =$} 
\mbox{$ i\gamma \cdot p\, A(p^2) + B(p^2)$}, this leads to the
correspondance \mbox{$\lambda_1^{\rm WI} = A$},
\mbox{$\lambda_2^{\rm WI} =$} \mbox{$ -A^\prime/2$}, and 
\mbox{$\lambda_3^{\rm WI} =$} \mbox{$ B^\prime$}, where 
\mbox{$f^\prime =$} \mbox{$ \partial f(p^2)/ \partial p^2$}.  
\begin{figure}[t] 
\vspace*{2em}
 \centerline{\includegraphics[width=0.45\textwidth]
                  {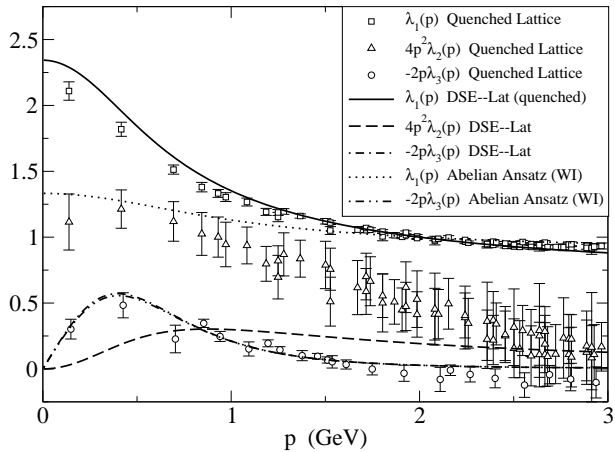}}

\caption{\label{fig:Lat_WI} The amplitudes of the dressed quark-gluon
vertex at zero gluon momentum and for quark current mass 
\mbox{$m(\mu=2~{\rm GeV}) =$} 60~MeV.  Quenched lattice 
data~\protect\cite{Skullerud:2003qu} is compared to the results of the
DSE-Lat model~\protect\cite{Bhagwat:2003vw}.  The Abelian Ansatz 
(Ward identity) is also shown except for 
$\lambda_2(p)$ which is almost identical to the DSE-Lat model.   }
\end{figure} 

Our nonperturbative model for the dressed quark-gluon vertex is defined by
extentions of  \Eqs{VertA} and (\ref{VertNA}) into dressed versions 
determined solely from an existing ladder-rainbow model DSE kernel that has
1-loop QCD renormalization group improvement.  Two DSE models are 
employed.  The first model (DSE-Lat)~\cite{Bhagwat:2003vw} represents 
a mapping of quenched lattice data for the  gluon 
propagator~\cite{Leinweber:1998uu} into a continuum ladder-rainbow  model
kernel having sufficient effective infrared vertex strength to
reproduce quenched lattice data for the quark propagator~\cite{Bowman:2002kn}. 
In this sense, it represents quenched dynamics.  The second 
(DSE-MT)~\cite{Maris:1999nt} provides a good one-parameter fit to a wide 
variety of light quark meson physics; in this sense it represents unquenched 
dynamics.    Both can be implemented through the
substitution \mbox{$g^2\, D_0(q^2) \to$} \mbox{${\cal G}(q^2)/q^2$}
in the ladder BSE kernel that appears in the integrand for
$\Gamma^{\rm A}_\sigma$, in \Eq{VertA}.  The bare quark propagators 
in \Eqs{VertA} and (\ref{VertNA}) are replaced
by solutions of the quark DSE in rainbow truncation using the appropriate 
kernel, namely, 
\begin{eqnarray}
\lefteqn{S(p)^{-1}\;=\;Z_2 \, i\,/\!\!\!p + Z_4 \, m(\mu)}
\nonumber\\ &&
        {} + \ssize{C_{\rm F}} \int^\Lambda_{p^\prime} \!
              \frac{{\cal G}(q^2)}{q^2}\, 
         T_{\mu\nu}(q) \,\gamma_\mu \, S(p^\prime) \, \gamma_\nu~,
\label{quarkdse}
\end{eqnarray}
where \mbox{$q=p-p^\prime$}.
For both models the effective coupling ${\cal G}(q^2)$ has a parameterized 
form in the infrared, while in the ultraviolet, the QCD factors from 1-loop
renormalization of the quark and gluon propagators and the pair of
quark-gluon vertices have been absorbed so that ${\cal G}(q^2)$ matches
$4\, \pi\,\alpha_s^{\rm 1-loop}(q^2)$~\cite{Maris:1997tm}.
The corresponding rainbow DSE solution reproduces the leading logarithmic
behavior of the quark mass function in the perturbative spacelike region.

A ladder-rainbow model kernel is generally not sufficient to specify
a dressed extension of $\Gamma^{\rm NA}_\sigma(p+k,p)$ from \Eq{VertNA}.
However at $k=0$, the expression in \Eq{VertNA0} for 
$\Gamma^{\rm NA}_\sigma(p,p)$ has combined the triple gluon vertex and 
the gluon propagators to produce a form that emphasizes the close 
connection to the ladder kernel and the self-energy integral.  
The same nonperturbative extension \mbox{$g^2\, D_0(q^2) \to$}
\mbox{${\cal G}(q^2)/q^2$} as used earlier
now suggests itself, and we use it.    Our justification
for this choice is one of simplicity;  no new parameters are introduced.
\begin{figure}[t] 
\vspace*{2em}
 \centerline{\includegraphics[width=0.45\textwidth]
                  {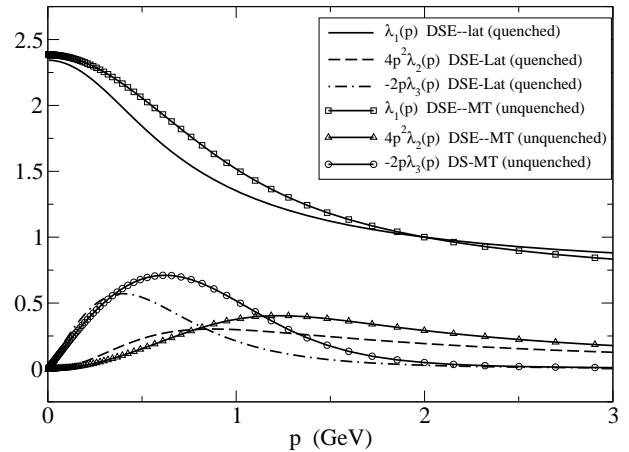}}

\caption{\label{fig:MT_Lat} The amplitudes of the dressed quark-gluon
vertex at zero gluon momentum, and for quark current mass 
\mbox{$m(\mu=2~{\rm GeV}) =$} 60~MeV, from two models: 
DSE-Lat~\protect\cite{Bhagwat:2003vw} and DSE-MT~\protect\cite{Maris:1999nt}
that relate to quenched and unquenched content respectively.}
\end{figure} 

\section{\label{sec:results} Results and discussion}

In \Fig{fig:Lat_WI} we display the DSE-Lat model results in a
dimensionless form for comparison with the (quenched) lattice 
data$^{\rm\footnotemark[2]}$\footnotetext[2]{We note 
that in \Ref{Skullerud:2003qu} both the lattice data, and the Abelian 
(Ward identity) Ansatz, for $\lambda_3(p)$ are presented as positive.  
These two sign errors have been acknowledged~\cite{Skull_PrivCom04}.}.  
The renormalization scale of the lattice data is \mbox{$\mu = 2$}~GeV
where \mbox{$\lambda_1(\mu) = 1$}, \mbox{$A(\mu) = 1$}.   We compare to the 
lattice data set for which \mbox{$m(\mu) = 60$}~MeV.  The same 
renormalization scale and conditions have been implemented for both DSE 
models$^{\rm\footnotemark[3]}$\footnotetext[3]{To facilitate change of the 
scale $\mu$, we have slightly 
modified both DSE kernels (both originally defined at fixed scale 
$\mu_0=19$~GeV) by including the additional kernel strength
factor $Z_2^2(\mu^2,\Lambda^2)/Z_2^2(\mu_0^2,\Lambda^2)$ recommended by 
Maris~\cite{MarisPrivCom}.  This does not alter results for observables. }.
For $\lambda_1$ and $\lambda_3$ we also compare with 
the Abelian Ansatz in which the amplitudes are obtained from the quark 
propagator through the Ward Identity, which is equivalent to the 
$k=0$ limit of  either the Ball-Chiu~\cite{Ball:1980ay} or 
Curtis-Pennington~\cite{Curtis:1990zs} Ansatz.  
Without parameter adjustment, the model reproduces the
lattice data for $\lambda_1$ and $\lambda_3$ quite well over the whole
momentum range for which data is available.
The Abelian Ansatz, while clearly inadequate for $\lambda_1$ below 1.5~GeV,
reproduces $\lambda_3$.    The present 
lattice data for $\lambda_2$ has large errors;  it suggests infrared 
strength that is seriously underestimated by the model.  (The Abelian 
Ansatz for $\lambda_2$ is very close to the DSE model and for reasons 
of clarity, is not displayed.)

The relative contributions to the vertex dressing made by 
$\Gamma^{\rm NA}_\sigma$ and $\Gamma^{\rm A}_\sigma$ are indicated by
the following amplitude ratios at \mbox{$p=0$}: 
\mbox{$\lambda^{\rm NA}_1/\lambda^{\rm A}_1 = -60$}, 
\mbox{$\lambda^{\rm NA}_2/\lambda^{\rm A}_2 = -14$},
and \mbox{$\lambda^{\rm NA}_3/\lambda^{\rm A}_3 = -12$}.   Thus the
non-Abelian term $\Gamma^{\rm NA}_\sigma$ dominates to a greater extent 
than what the ratio of color factors ($-9$) would suggest; it also 
distributes its infrared strength to favor $\lambda_1$ more so than does
$\Gamma^{\rm A}_\sigma$.    Since the momentum-dependent shapes of the 
$\lambda^{\rm NA}_i(p)$ and $\lambda^{\rm A}_i(p)$  are quite similar, 
the present model results could be summarized
quite effectively by ignoring $\Gamma^{\rm A}_\sigma$ and scaling
$\Gamma^{\rm NA}_\sigma$ up by about 10\%.   

Due to the definition of the two DSE models,  their comparison  in 
\Fig{fig:MT_Lat} provides an estimate of the effects of the quenched 
approximation.  The effects are moderate within the present DSE 
model framework.   \Fig{fig:MT_Lat} also suggests  that a model including 
the four gluon vertex as well as the two 
diagrams of \Fig{fig:2vertdiags} should be considered, especially for amplitude
$\lambda_2$. 
The question of the importance of the iterations of the diagrams of 
\Fig{fig:2vertdiags} also arises.   We have estimated such effects
by iteration to all orders based on the ladder-rainbow kernel.   This amounts
to solution of a ladder Bethe-Salpeter integral equation in which the 
inhomogeneity is our dressed extension of \mbox{$Z_{\rm 1F}\,\gamma_\sigma + 
\Gamma_\sigma^{\rm NA}(p,p)$} and the kernel term is the dressed extension of
$\Gamma_\sigma^{\rm A}(p,p)$ with the internal $\gamma_\sigma$ replaced by
$\Gamma_\sigma(q,q)$.   This generates very little change---significantly
less than the quenching effect evident in \Fig{fig:MT_Lat}.  This is due 
to the small color factor of the kernel term.  We have not explored the 
consequences of using the 
dressed vertex self-consistently for the internal quark-gluon vertices of 
$\Gamma_\sigma^{\rm NA}$ in \Fig{fig:2vertdiags}-NA.
  
The nonperturbative Ansatz we have applied to \Eq{VertNA0} is equivalent to
the use of an effective dressed triple-gluon vertex $\Gamma^{3g}_{\mu \nu \sigma}$
satisfying \Eqs{3GSTI} and (\ref{3GSTI0}) with
the substitution \mbox{$D_0(q^2) \to$} \mbox{$[{\cal G}(q^2)/g^2(\mu^2)]/q^2$}.
Some perturbative studies of $\Gamma^{3g}_{\mu \nu \sigma}$ have been made at 
1-loop~\cite{Davydychev:1996pb,Davydychev:2001uj} but they provide no guidance 
for extension to  infrared scales.   The nonperturbative Ans\"atze for 
$\Gamma^{3g}_{\mu \nu \sigma}$ suggested in \Refs{Alkofer:2000wg} and 
\cite{Fischer:2002eq} for use within truncated gluon-ghost-quark DSEs require
explicit models for the ghost dressing function and the ghost-gluon 
vertex that appear in the STI for $\Gamma^{3g}_{\mu \nu \sigma}$.
Such considerations are beyond the scope of the present work; they would 
entail additional parameters that are not warranted at this stage.    
Recent work on a model of the quark-gluon vertex using such an Ansatz for
$\Gamma^{3g}_{\mu \nu \sigma}$ has produced results similar to the present work,
except that the \mbox{$m(\mu) = 115$}~MeV case is considered~\cite{FischerAdel04}.
Evidently the detailed infrared structure of $\Gamma^{3g}_{\mu \nu \sigma}$ is 
not crucial to present considerations.

\section*{Acknowledgments} 
\noindent
The authors would like to thank R.~Alkofer, C.~D.~Roberts, A.~{K\i z\i lers\"u},
and J.~I.~Skullerud,  for useful discussions.  We are grateful to J.~I.~Skullerud
for providing the lattice-QCD results.  This work has 
been partially supported by  NSF grants no. PHY-0301190 and no. INT-0129236.  


\bibliography{refsPM,refsPCT,refsCDR,refs}


\end{document}